\documentclass[12pt,preprint]{aastex}

\newcommand{\rmd}{ {\ \mathrm d} }
\renewcommand{\d}{\mbox{d}}

\newcommand{\mi}[1]{\mbox{\boldmath$#1$}}

\shorttitle{Acoustic travel time shifts}
\shortauthors{Parchevsky et al.}

\begin{document}


\title{Verification of the travel time measurement technique and the helioseismic inversion procedure for sound speed using artificial data}


\author{K.V.~Parchevsky, J.~Zhao, T.~Hartlep, and A.G.~Kosovichev}
\affil{Stanford University, HEPL, Stanford CA 94305}
\email{kparchevsky@solar.stanford.edu}



\begin{abstract}
We performed 3D numerical simulations of the solar surface wave field for the quiet Sun and for three models with different localized sound-speed variations in the interior with: (i) deep, (ii) shallow, and (iii) two-layer structures. We used simulated data generated by two different codes which use the same standard solar model as a background model, but utilize two different integration techniques and use different models of stochastic wave excitation. Acoustic travel times were measured from all data sets using the time-distance helioseismology technique and compared with the ray theory predictions, frequently used for helioseismic travel-time inversions. It is found that the measured travel-time shifts agree well with the ray theory in both cases with and without phase-speed filtering for the shallow and deep perturbations. This testing verifies the whole measuring-filtering-inversion procedure for sound-speed anomalies inside the Sun. It is shown, that the phase-speed filtering, frequently used to improve the signal-to-noise ratio does not introduce significant systematic errors. Results of the sound-speed inversion procedure show good agreement with the background sound-speed profiles in all cases. Due to its smoothing nature, the inversion procedure overestimates sound speed variations in areas with sharp gradients of the sound-speed profile.
\end{abstract}





\section{Introduction}
     \label{S-Introduction}
Both time-distance helioseismology \citep{duv93} and phase-sensitive helioseismic holography \citep{bra00} measure acoustic travel-times or acoustic wave phase shifts to infer solar interior properties such as sound-speed perturbations and flow fields. For time-distance helioseismology inversions, the measured travel-time anomalies associated with sunspots and active regions are often interpreted as caused by solar subsurface sound-speed perturbations (e.g.~\citet{kos00}; \citet{cou04}). The travel-time inversions for the sound-speed perturbations are usually done by using a variational formulation of the acoustic ray theory \citep{kos97} or the first Born approximation \citep{bir00,bir04} and provide similar results for sunspot data \citep{cou04}.

It is important to verify and calibrate the standard helioseismic inversion procedure, which can be done by analyzing artificial data obtained from numerical simulations. We apply the time-distance measurement and helioseismic inversion procedures to numerically simulated stochastic velocity wave fields on the Sun calculated using different simulation codes. Both codes solve the linearized Euler equations. The Cartesian code \citep{Parchevsky2007a} uses a finite-difference scheme with impulsive localized wave sources randomly distributed at same depth and random in time. Another code \citep{har05} is spherical and utilizes a combination of a spectral method in angular coordinates and B-splines in radial direction. Wave sources used in this code are distributed and defined in the Fourier space as a combination of spherical harmonics with random amplitude and phase and a mean acoustic power spectrum similar to the solar one.

Our goal is to examine how well the time-distance measurements and sound-speed inversion results based on acoustic ray-path approximation agree with the travel-time shifts expected from the acoustic ray theory and sound-speed variations of the background model. Similar testing for the travel times measured by the acoustic holography method and sound-speed inversions based on the Born approximation has been published by \citet{bir11}. We introduce the models and numerical simulations in \S2. In \S3 we describe details of the time-distance measurement techniques with and without phase-speed filtering. In \S4 we present results of sound-speed inversions for different profiles of the background sound speed variations using the artificial data, obtained by different codes. Conclusions follow in \S5.

\section{Artificial data}
\subsection{Simulation of local wavefield}
We use the linearized code for simulations of wave propagation in the Cartesian geometry in the Sun, described by \citet{Parchevsky2007a}. A finite difference scheme, adapted to the non-uniform vertical grid, is used for the spatial discretization. A strong-stability-preserving Runge-Kutta algorithm is used for time advancing \citep{Shu2002}. We simulated the wave field generated by localized sources with amplitude $A$, added in the right hand side of the momentum equation as a vertical force
\begin{equation}\label{Eq:SrcXYZ}
S_z(r,t)= A G(r)F(t),
\end{equation}
randomly distributed in a shallow layer 100 km below the photosphere.
The spatial $G(r)$ and temporal $F(t)$ behavior of the wave source is modeled by functions
\begin{equation}\label{Eq:SrcGr}
G(r) =
\left(1-\frac{(z-z_{src})^2}{H_{src}^2}\right)^2\left(1-\frac{(x-x_{src})^2+
(y-y_{src})^2}{R_{src}^2}\right)^2,
\end{equation}
\begin{equation}\label{Eq:SrcFr}
F(t)= \left(1-2\tau^2\right)e^{-\tau^2},
\end{equation}
where $R_{src}$ and $H_{src}$ are the source radius and height respectively, $x_{src}$, $y_{src}$, and $z_{src}$ are the coordinates of the source origin. Parameter $\tau$ is given by equation
\begin{equation}
\tau=\frac{\omega_0 (t-t_0)}{2} - \pi, \qquad t_0\leq t\leq
t_0+\frac{4\pi}{\omega},
\end{equation}
where $\omega_0$ is the central source frequency, $t_0$ is the moment of the source initiation. In our simulations we chose $H_{src}$ = 0.1 Mm, $R_{src}$ = 0.3 Mm. This source model provides the wave spectrum, which closely resembles the solar spectrum. It has a peak near the central frequency $\omega_0$ and spreads over a broad frequency interval. The source spectrum is \citep{Parchevsky2008}:
\begin{equation}
|\hat{F}(\omega)| \equiv \left|\int_{-\infty}^\infty F(t)e^{-i\omega t} \rmd t \right|
=4\sqrt{\pi}\;\frac{\omega^2}{\omega_0^3}\;e^{-\frac{\omega^2}{\omega_0^2}}.
\end{equation}
The acoustic wave field, generated by this code with described distribution of wave sources, has a power spectrum that is very similar to the spectrum obtained from SOHO/MDI high-resolution observations (Fig. \ref{Fig:k-w}).

Great efforts were made to develop the top boundary condition which mimics non-reflective properties of the solar atmosphere. Waves with frequencies below the acoustic cut-off frequency are reflected from the sub-photospheric layer with strong density and pressure gradients (e.g.~\citet{Gough1993}). The waves with frequencies higher than the acoustic cut-off frequency, pass through and propagate to the chromosphere and escape to the corona. For simulation of this property, we set a wave absorbing layer, based on Perfectly Matched Layer (PML) technique, in low chromosphere, about 500 km above the photosphere. The waves with the frequencies higher than the acoustic cut-off frequency are absorbed by this layer and do not pollute the computational domain. The higher layers are not considered because they are not essential for the current helioseismology study. A similar absorbing layer was set at the bottom of the domain. The lateral boundary conditions were chosen periodic.

We built a background plain-parallel model of the Sun by smoothly joining the model of the chromosphere (VALC) of \citet{Vernazza1981} with the standard solar model S of \citet{Christensen-Dalsgaard1996}. To eliminate the convective instability in the subsurface superadiabatic region, the background model was stabilized against convection by setting the negative entropy gradient in the entropy equation to zero (see \citet{Parchevsky2007a}).

\subsection{Simulations of whole-Sun oscillations}

Simulations of the global full-Sun acoustic wave field are preformed with a code developed by~\citet{har05}. This code solves the linearized propagation of helioseismic waves throughout the entire solar interior, and has been used in various studies to simulate the effects of localized sound speed perturbations, e.g., for testing helioseismic far-side imaging by simulating the effects of model sunspots on the acoustic field~\citep{2008ApJ...689.1373H,2009SoPh..258..181I}, for validating time-distance helioseismic measurements of tachocline perturbations~\citep{2009ApJ...702.1150Z}, and for studying the effect of localized subsurface perturbations \citep{2011SoPh..268..321H}. It models solar acoustic oscillations in a spherical domain by using the Euler equations linearized around a stationary, non-rotating, background state, and uses several approximations. In particular, perturbations of the gravitational potential have been neglected, and the adiabatic approximation has been used. Also, the entropy gradient of the background model has been neglected in order to make the linearized equations convectively stable. The calculations show that this assumption does not significantly change the propagation properties of acoustic waves including their frequencies, except for the acoustic cut-off frequency, which is slightly reduced. Non-reflecting boundary conditions are used at the upper boundary by means of an absorbing buffer layer with a damping coefficient that is zero in the interior and increases smoothly into the buffer layer.

A random function was added in the mass continuity equation to mimic the excitation of acoustic sources, which is a non-linear process not included the in linearized equations. This function is non-zero only close below the solar surface. It is random in latitude and longitude, and in time, but with a prescribed temporal spectrum approximating the solar acoustic power spectrum.

The equations are solved using a pseudo-spectral method with a spherical harmonic discretization in the angular directions, and 4th order B-splines \citep{Loulou97,1999JCoPh.151..757K} for the radial direction. 2/3-dealiasing is used. The radial resolution of the B-spline method is varied proportionally to the local speed of sound so that the method's resolution in greater near the surface (where the sound speed is small), and coarser in the deep interior (where the sound speed is large). The simulation in this study resolves spherical harmonics of angular degree between 0 to 170, and uses 350 B-splines in the radial direction. A staggered \citet{1966ITAP...14..302Y} scheme is used for time integration, with a time step of 1 seconds.

There are two main differences between the Cartesian (CC) and the spherical (SC) codes: (i) CC uses finite difference spatial scheme, SC uses decomposition of the wave field on spherical harmonics in the horizontal direction, (ii) CC uses vertical dipole sources localized in space and time and defined in the configurational space, SC code uses sources defined in the Fourier space as a composition of spherical harmonics with random parameters. So, in the configurational space, sources are not localized, but rather smeared over the whole solar surface. These two different approaches to the simulation of the wavefield in the Sun produce the acoustic wave fields with similar parameters.

\section{Travel-time measurement technique}
Travel-time measurements are based on fitting of the cross-covariance function
\begin{equation}\label{Eq:CrossCorr}
C(\mi{r}_1,\mi{r}_2,t)=\frac{1}{T}\int_0^T\phi(\mi{r}_1,t')\phi(\mi{r}_2,t'+t)\d t'
\end{equation}
between two source and destination points with horizontal radius-vectors $\mi{r}_1$ and $\mi{r}_2$, where  $T$ is the total time of observation, $t$ is the time lag between two signals, and $\phi$ is the vertical (z-component) of the velocity perturbation in the wave. An $f$-mode filter is applied to $\phi(\mi{r}_1,t)$ and $\phi(\mi{r}_2,t)$ to filter out the $f$-mode. Then, to increase the signal-to-noise ratio, the point-to-point cross-covariance functions are averaged over the annulus with internal and external radii $\Delta_{min}, \Delta_{max}$, centered at the source. To do this, the annulus is divided along the radius into pixel-wide sub-annuli and point-to-annulus cross-covariance functions are calculated for each sub-annulus by averaging the point-to-point cross-covariance functions along each sub-annulus.  Then, the individual point-to-annulus cross-covariance functions are shifted in time in such a way, that they are centered on the specific reference time, which corresponds to the travel time between the source and the sub-annulus with radius $1/2(\Delta_{min}+\Delta_{max})$. To produce the final point-to-annulus cross-covariance function, the time shifted cross-covariance functions for individual sub-annuli are averaged.

\begin{figure}
\includegraphics[width=0.9\textwidth]{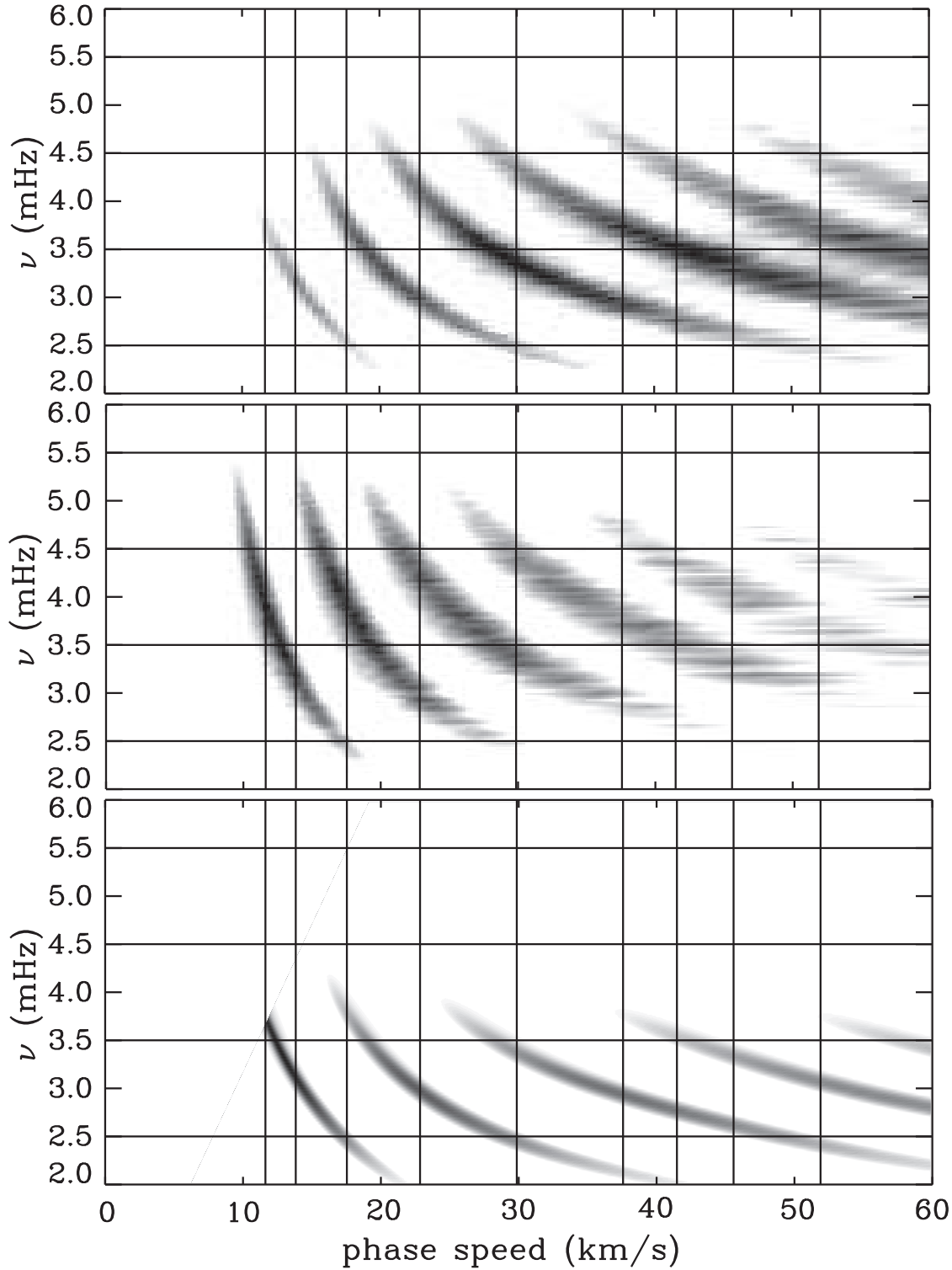}
\caption{Oscillation power spectra from high-resolution SOHO/MDI observations ({\sl top}), and from the two numerical simulations of the quiet Sun in the Cartesian geometry ({\sl middle}), and in the spherical geometry ({\sl bottom}) as functions of the phase speed, $\omega/k$, and cyclic frequency, $\nu=\omega/2\pi$. The leftmost ridge corresponds to the $f$-mode (surface gravity wave), and other ridges represent acoustic ($p$) modes. The vertical lines indicate where the phase-speed filter allows the passage of a half of the power, and the horizontal lines show the frequency bands that are used for the frequency dependent studies. }\label{Fig:k-w}
\end{figure}

\citet{kos97} proposed to fit the cross-covariance function by the Gabor's wavelet
\begin{equation}\label{Eq:GaborsWavelet}
G(A,\omega_0,\delta\omega,\tau_p,\tau_g;t)=A\cos(\omega_0(t-\tau_p))
\exp\left(\frac{\delta\omega^2}{4}(t-\tau_g)^2\right),
\end{equation}
where $A$ is the amplitude, $\omega_0$ is the central frequency, $\delta\omega$ is the frequency width, $\tau_p$ is the phase travel-time, and $\tau_g$ is the group travel-time. Equation (\ref{Eq:GaborsWavelet}) can be considered as a definition of the phase- and group travel times. This definition is not unique. \citet{Gizon2002} proposed a definition of the wave packet travel time based on geophysical studies. Later, \citet{Gizon2004} proposed another definition of the travel-time linear in the cross-covariance. This approach was described by \citet{bir09,bir11}

\begin{figure}
\includegraphics[width=0.6\textwidth]{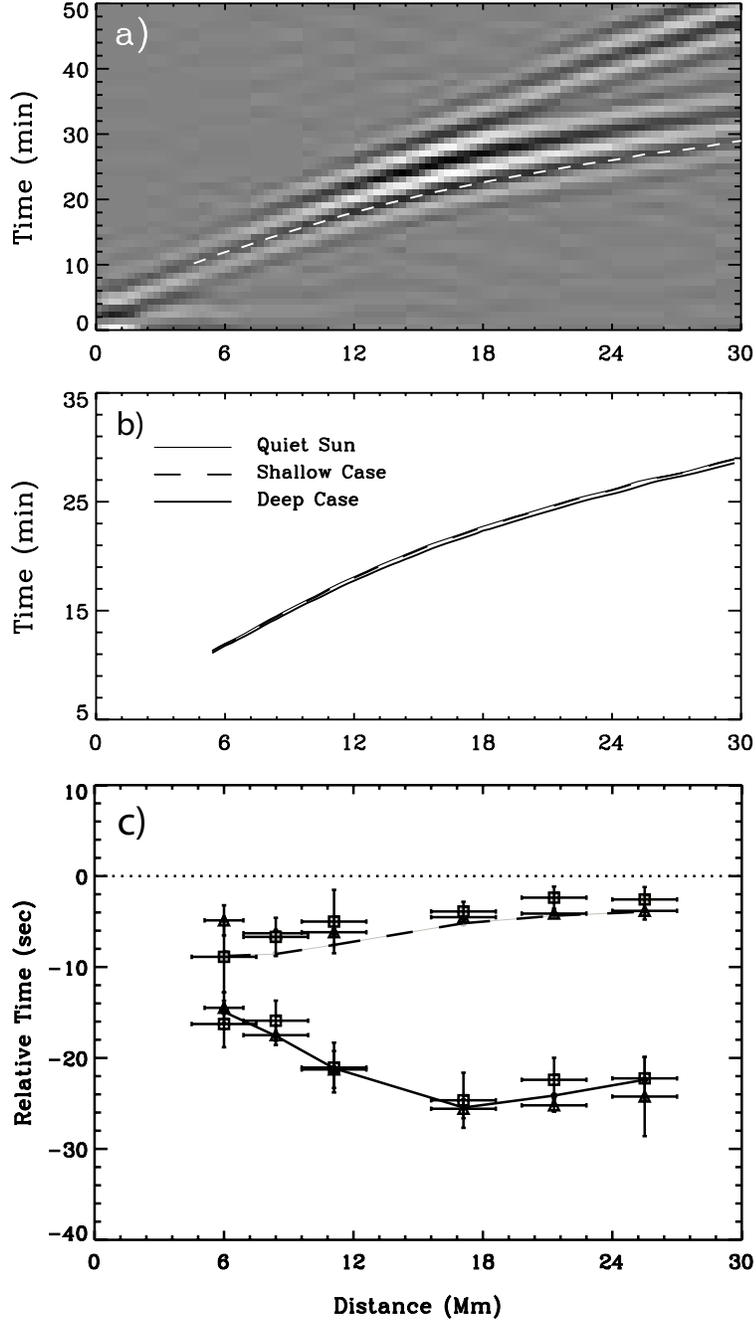}
\caption{(a) Example of the cross-covariance function (time-distance diagram) from the numerical simulations, using the Cartesian geometry, averaged over the selected central area. White dashed curve shows the Gabor's wavelet fitted phase-travel times. (b) Comparison of the phase travel times measured from the three simulation models: quiet Sun (thin solid curve), cases with shallow (dashed curve) and deep (solid curve) perturbations. (c) Travel time shifts obtained without phase-speed filtering (triangles) and with phase-speed filtering (squares) from the simulation data relative to the quiet Sun. Dashed and solid curves represent time shifts calculated from the acoustic ray theory for the shallow and deep models respectively.}\label{Fig:TD}
\end{figure}

Usually, before the averaging, $\phi(\mi{r},t)$ is filtered in the Fourier domain using a phase-speed filter with the Gaussian profile \citep{Duvall1997}. Such filtering selects only waves which travel with phase speed in a given range, or, in other words, waves which travel-time between the source and destination lies in some range which depends on the width of the pase-speed filter. This further improves the signal-to-noise ratio. The central phase-speed and the width of the phase-speed filter depend on the internal and external radii $\Delta_{min}, \Delta_{max}$ of the annulus and must be carefully tuned out. Here we perform the time-distance analysis of the simulation data with and without the phase-speed filtering procedure in order to test the accuracy of the filtering.

\begin{figure}
\includegraphics[width=\textwidth]{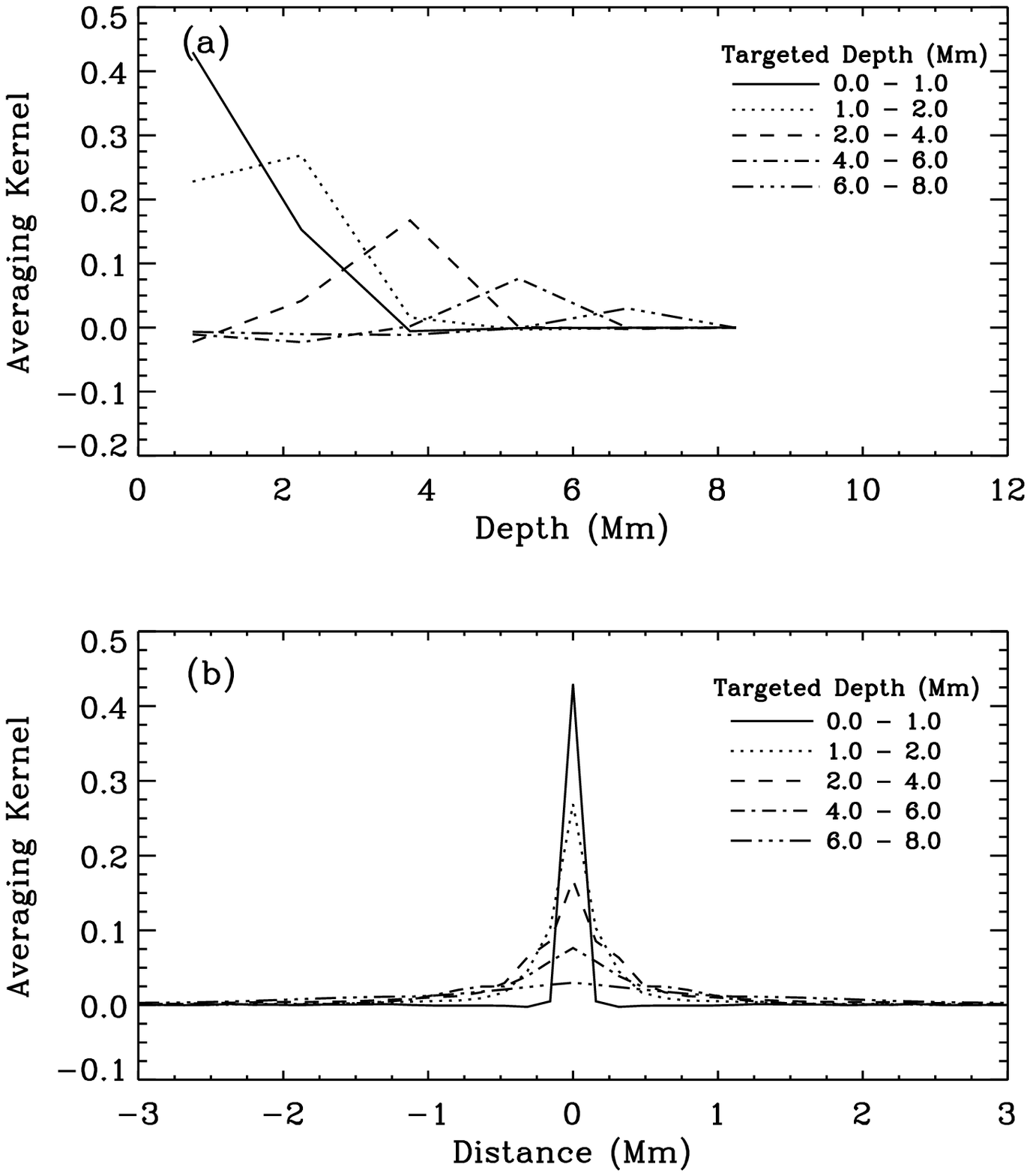}
\caption{Vertical (top panel) and horizontal (bottom panel) slices of the averaging kernels used for the helioseismic inversions of simulated data for the shallow, deep and two-layer models, described in the text. The solid, dotted, dashed, dash-dotted, and dash-dot-dotted curves correspond to the depth ranges of 0-1, 1-2, 2-4, 4-6, 6-8 Mm respectively.} \label{Fig:AvrKernels}
\end{figure}
\begin{figure}
\includegraphics[width=\textwidth]{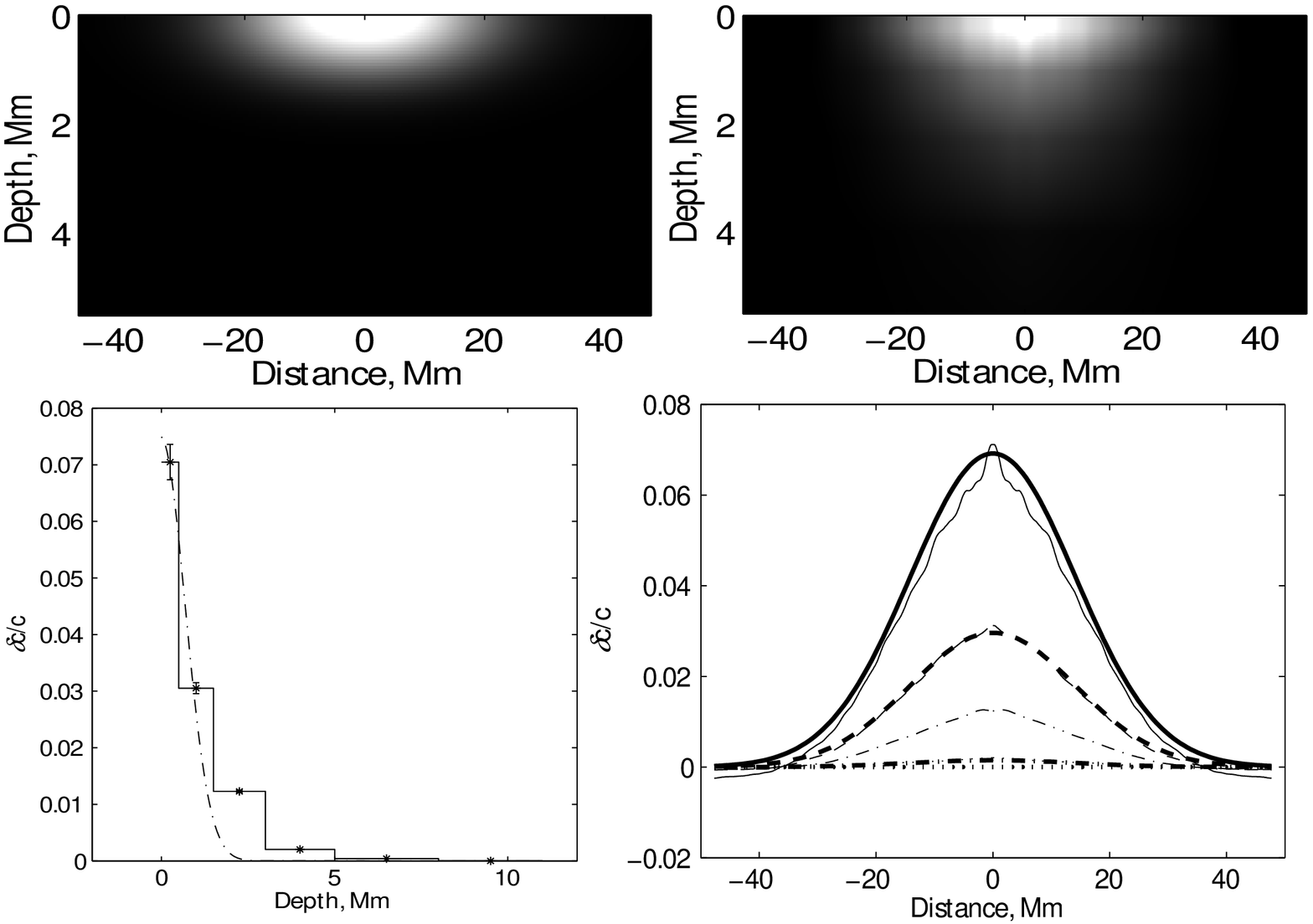}
\caption{Model and reconstructed sound speed profiles for the shallow model. The top left panel shows vertical cut of the sound speed perturbation, used as the background state in the shallow model. The top right panel shows the sound-speed perturbation obtained by inversion of travel times. The bottom left panel shows the sound-speed perturbation at the axis of the shallow model as a function of depth. The dash dotted curve represents the model profile given by equation (\ref{Eq:SoundSpeedVariation}), the solid step curve shows the mean reconstructed profile. The error bars represent the RMS deviations of the reconstructed sound speed perturbation in horizontal plane. The bottom right panel represents the 1D horizontal profiles of the sound speed perturbation averaged over depth ranges 0-0.5, 1.5-3, 5-8, and 8-11 Mm, shown by solid, dashed, dot-dashed, and dotted curves respectively. The thick curves show profiles in the background model, the thin curves represent inversion results. } \label{Fig:ShallowModel}
\end{figure}

\section{Results and discussion} \label{S-Results}
In this section we present results of numerical simulations, time-distance analysis of the artificial data, and results of the helioseismic inversion of the sound speed profiles for different models of the sub-photospheric sound speed perturbations. To simulate local subsurface sound-speed structures, two types of axially symmetric perturbations of the background sound speed were introduced: constant-sign perturbation (two models with different parameters) and two-layer perturbation (one model),  where the background sound speed perturbation changes sign at some depth. The spatial distribution of constant-sign sound-speed perturbations is given by equation
\begin{equation}\label{Eq:SoundSpeedVariation}
\delta c^2(r,z)/c_0^2(z) = \epsilon\exp(-r^2/R_{spt}^2)\exp(-z^2/D_{spt}^2),
\end{equation}
where $r$ is the horizontal distance from the symmetry axis of the perturbation, $z$ is the depth relative to the photosphere, $c_0(z)$ is the sound speed in the quiet Sun model, $R_{spt}$ is the horizontal scale of the sound speed perturbation, $D_{spt}$ is the depth scale, and $\epsilon$ is the fractional amplitude of the perturbation in the sound speed frame. We adopted two different sets of parameters for the constant-sign perturbation from \citet{bir09}: "shallow" case (a) with $D=1$~Mm, $R=20$~Mm, and $\epsilon=0.15$, and a "deep" case (b) with $D=10$~Mm, $R=20$~Mm, and $\epsilon=0.1$. Sound-speed perturbations in both "deep" and "shallow" cases are positive. The two-layer sound-speed perturbation is negative in a shallow region below the photosphere and positive for depths greater than 3 Mm. The vertical profile of the sound speed in the two-layer model was chosen to resemble the initial helioseismic sound-speed inversion of a sunspot \citep{kos00}. The horizontal profiles in all three cases were chosen Gaussian with $R=20$~Mm. In addition to these simulations with three different models of the background sound-speed perturbation, we performed simulations with the horizontally uniform "quiet" solar model. Locations, amplitudes, frequencies, and moments of initiation of the wave sources in simulations with the quiet solar model and all three perturbed models were the same.

We performed 3D numerical simulations in the Cartesian geometry for all three models of sound-speed perturbations: (i) "deep", (ii) "shallow", and (iii) two-layer perturbation. Simulations were carried out in a box of size $96\times 96\times 36.8$ Mm$^3$ using grid size of $640\times640\times92$. The horizontal grid is uniform ($\Delta x=\Delta y =150$~km), and the vertical grid is non-uniform with spacing $\Delta z=50$~km above the photosphere and gradually increasing to $\Delta z = 1.76$~Mm at the bottom of the of the computational domain. The time step was chosen to be 0.5 s to comply the CFL stability condition of the explicit numerical scheme. In all three models localized stochastic sources of the vertical component of force with random amplitudes and frequencies were uniformly distributed at random locations at the depth of 100~km below the photosphere. We use 512-min time sequences with a 1-min cadence of the photospheric vertical velocity in a $15\times 15$ Mm$^2$ area for the time-distance analysis.

\begin{figure}
\includegraphics[width=\textwidth]{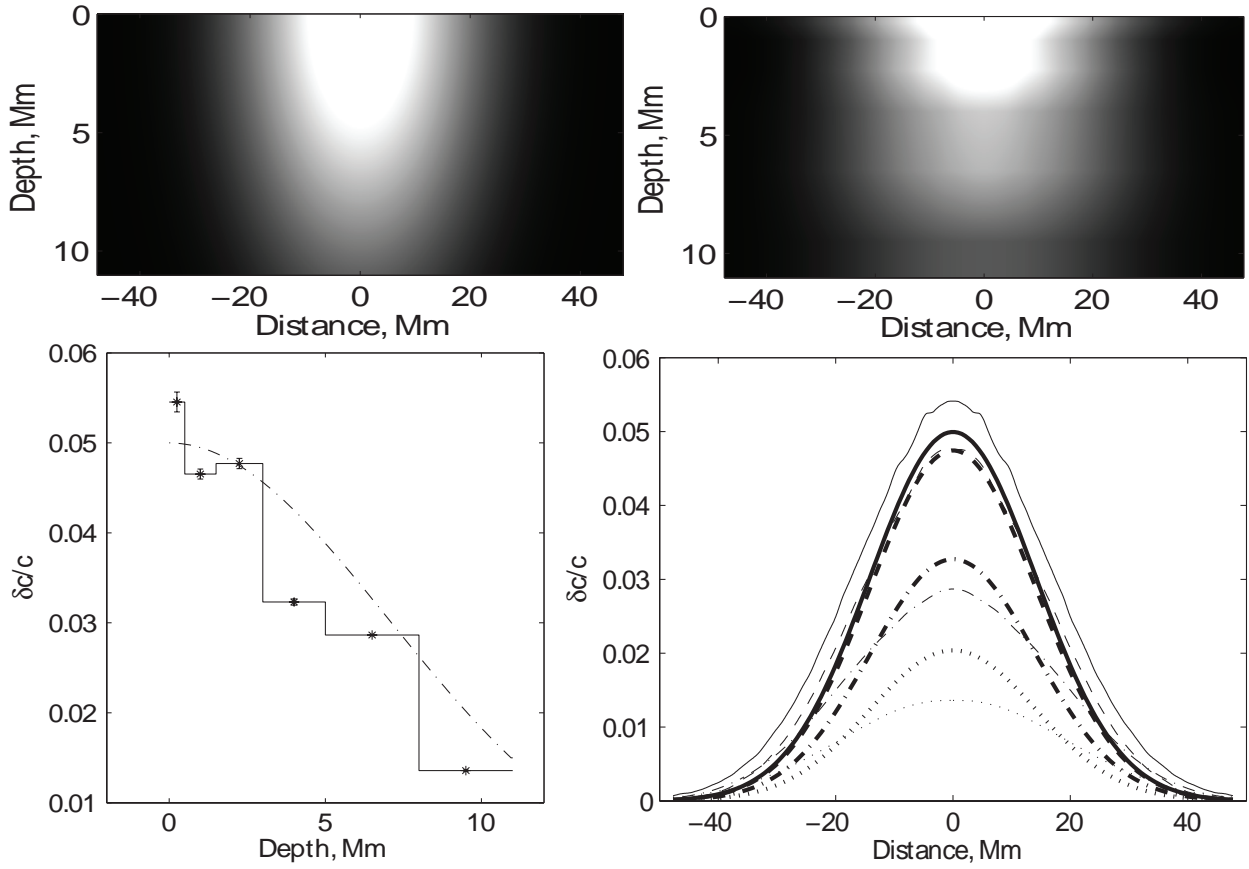}
\caption{Model and reconstructed sound speed profiles for the deep model. Position of panels and curve notations are the same as in Figure~\ref{Fig:ShallowModel}} \label{Fig:DeepModel}
\end{figure}

The oscillation power spectra obtained from the MDI and the simulations data are plotted in Figure~\ref{Fig:k-w} as functions of the oscillation cyclic frequency, $\nu=\omega/2\pi$, and the phase speed, $\omega/k$, where $k$ is the wavenumber, $\omega$ is the angular frequency. Position of the $f$- and $p$-mode ridges in all three spectra are nearly identical. The ridge width in the MDI power spectrum and power spectrum obtained from simulations in the Cartesian geometry is the same. The power distribution of the simulated wavefield in the Cartesian geometry is a bit stronger in the high-frequency and low phase-speed region, and a bit weaker in the low-frequency and high-phase-speed region. The ridge width of the power spectrum obtained from the simulations in the spherical geometry is smaller than in the MDI spectrum (probably due to lower dissipation in the sponge region). The simulated wavefield obtained from the spherical simulations exhibits excess of power at frequencies below 2.5~mHz and lack of power at frequencies above 4~mHz due to the reduced acoustic cut-off frequency caused by the non-inclusion of the entropy gradient. However, from our experiments of applying frequency filters, we find that these differences between the observed and simulated power spectra do not significantly alter the travel-time measurements in this work. The differences in power spectra from simulations in the Cartesian and spherical geometries are due to the different formulations of wave source excitation used in these simulations.

The cross-covariance function (time-distance diagram), obtained from the numerical simulations with the quiet-Sun background model is shown in panel (a) of Figure \ref{Fig:TD}. The cross-covariance function was calculated without phase-speed filtering which was introduced by \citet{Duvall1997} to improve the signal-to-noise ratio for the function. The fitted phase travel times for three simulations: quiet-Sun, and shallow and deep perturbation background models are shown in panel (b). A comparison of the phase travel times clearly shows that for the cases with both shallow and deep perturbations, the measured travel times are shorter than the travel times for the quiet Sun model, calculated from the acoustic ray theory, as expected for the positive background sound-speed perturbations. The phase travel time shifts measured with and without phase-speed filtering are shown in panel (c). The horizontal error bars indicate the annulus radius ranges used in this analysis. The vertical bars are formal fitting errors. The filtering parameters are adopted from \citep{cou06} and illustrated by vertical lines in Figure~\ref{Fig:k-w}. Because of the limited size of the simulation box only the first few filters are applied in our study. Essentially the same filter parameters were used by \citet{bir09}. Clearly, there is a good agreement between the measurements with and without the phase-speed filtering. At most points, the theoretical values fall within the scale of error bars of the measured travel times. This analysis provides an important validation for the basic technique of time-distance helioseismology. Of course, the acoustic ray theory has well-known limitations. It is expected that the theoretical description of acoustic travel times can be improved by accounting for the finite wavelength effects, e.g. by using the Born approximation \citep{bir00,bir04,bir11}.

\begin{figure}
\includegraphics[width=\textwidth]{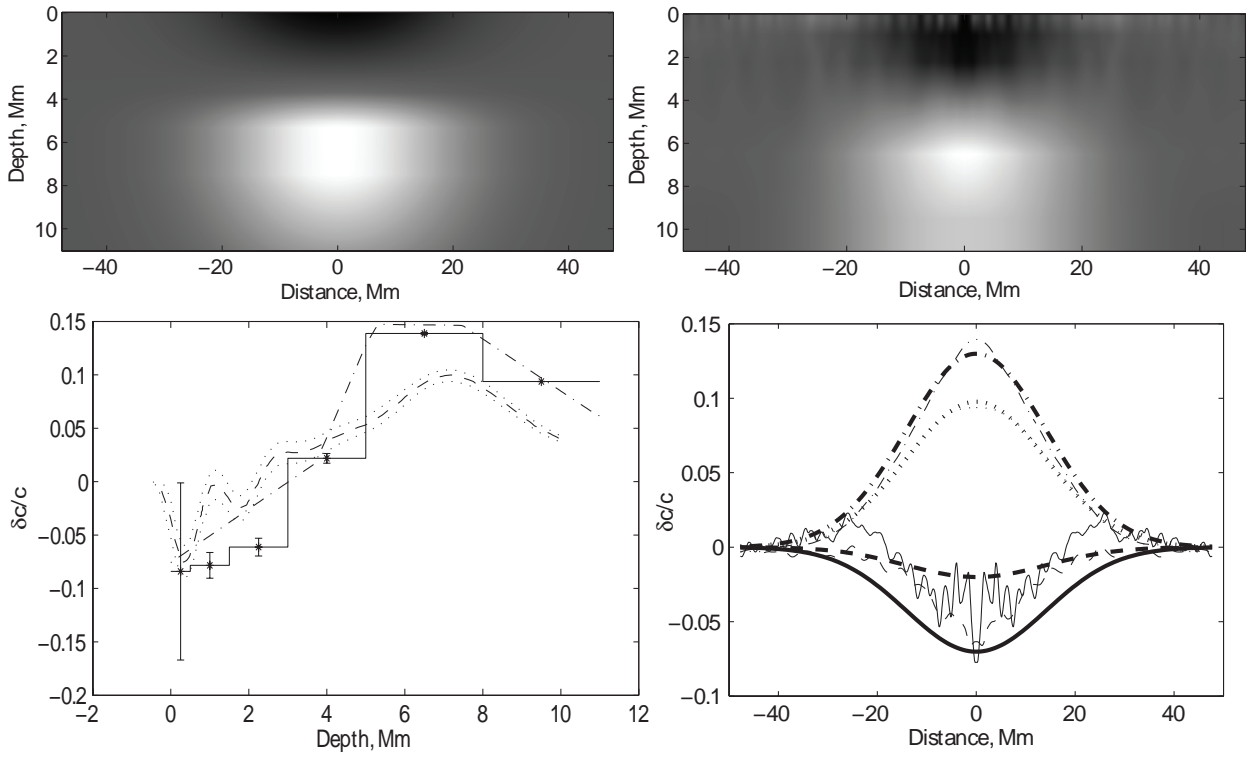}
\caption{Model and reconstructed sound speed profiles for the two-layer model. Position of panels and curve notations are the same as in Figure~\ref{Fig:ShallowModel}.} \label{Fig:TwoLayerModel}
\end{figure}

The sound-speed perturbations for the shallow model are presented in Figure~\ref{Fig:ShallowModel}. The top panels represent the background model sound-speed perturbation (left) and sound-speed perturbation reconstructed from the travel times by the helioseismic inversion (right). Bottom left and right panels represent the vertical and horizontal profiles of the sound-speed perturbation respectively. The reconstructed and model profiles show very good agreement above 2~Mm. The inversion results overestimate the sound-speed variations in the depth range from 1.5~Mm to 3.0~Mm, because the boundary of the sound-speed perturbation is located near the 1.5~Mm. The reconstructed sound-speed profile is smoothed, which causes a slight increase of the perturbation depth, which one can expect from the inversion procedure. Below 3.0~Mm both model and reconstructed sound speed perturbation profiles are zero. Similar results for the deep perturbation model are shown in Figure~\ref{Fig:DeepModel}. As for the shallow case, the reconstructed profiles of the sound-speed perturbation show good agreement with the sound-speed perturbation profiles in the background model except the depth region between 8~Mm and 11~Mm, which is close to the bottom boundary of the sound speed perturbation.

Results of reconstruction of the sound-speed perturbation in two-layer model are shown in Figure~\ref{Fig:TwoLayerModel}. Description of panels is the same as in Figure~\ref{Fig:ShallowModel}. The left and right top panels represent maps of the model and reconstructed sound speed perturbations respectively, the bottom left and right panels show vertical and horizontal profiles of the sound-speed perturbation. For comparison, we present inversion results obtained from the same simulation data by \citet{bir11}. They are shown by the dashed curve in the left bottom panel. Dotted curves show the error range. The reconstructed profiles are in good agreement with the model profiles. The depth of the sign inversion of the sound speed perturbation is located at the correct depth. Reconstructed amplitudes of the perturbation are close to the model ones, though the inversion slightly overestimates the perturbation amplitude in dept range from 1.5~Mm to 3.0~Mm, and are noisy in the shallow subphotospheric region (depth range from 0~Mm to 0.5~Mm).

We also performed numerical simulation of the wave field in the two-layer model of the sound speed perturbation in the spherical geometry using the spectral code of \citet{har05}. It is important to test the measurement and inversion procedures on different sets of artificial data, generated by the different codes. The spherical code allows to perform the testing for deeper perturbations than the Cartesian code, the depth of the perturbation was increased up to 20~Mm, when the depth of the original perturbation is 10~Mm.

\begin{figure}
\includegraphics[width=\textwidth]{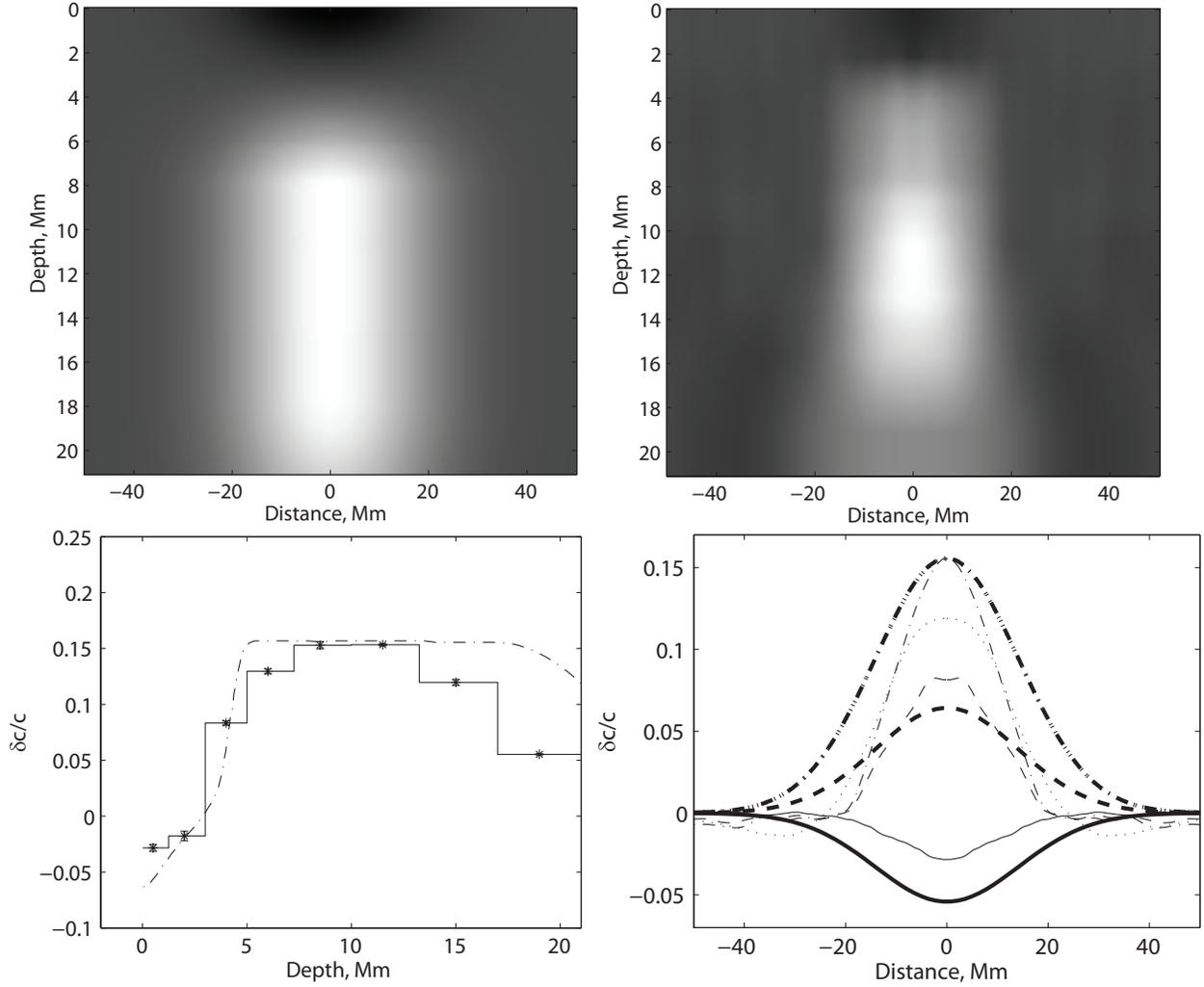}
\caption{Reconstruction of the sound speed perturbation for the two-layer model calculated from simulations in spherical geometry with the spectral code. Panel description is the same as in Figure~\ref{Fig:ShallowModel}. The sound speed perturbation used as a background model in spherical simulations is shown by the dash-dotted curve in bottom left panel. The solid, dashed, dot-dashed, and dotted curve in the right bottom panel represent horizontal profiles of the sound-speed perturbations averaged over depth ranges 0-1, 3-5, 7-10, 13-17 Mm respectively.} \label{Fig:ThomasSim}
\end{figure}

Results of the sound speed inversion in the two-layer model from simulations in spherical geometry with the spectral code are presented in Figure~\ref{Fig:ThomasSim}. We see good agreement between the background and reconstructed profiles of the sound speed perturbation, though the transition between negative and positive perturbations in the inversion results is not as steep as in the background model due to smoothing properties of the inversion procedure.

\section{Conclusions}
By using numerical simulations of stochastically excited solar oscillations we have tested the measurement and inversion procedures of time-distance helioseismology, based on the Gabor wavelet fitting to the cross-covariance function and ray-path sensitivity kernels. We have measured the travel time shifts caused by the subsurface sound-speed perturbations with and without phase-speed filtering for three different background distribution of the sound speed perturbation. We used two sets of simulated data generated by two different codes. One code uses finite difference spatial scheme in the Cartesian geometry and dipole wave sources localized in space and time. Another code uses spherical harmonic expansion in spherical geometry and distributed sources defined in the Fourier space. Simulations have been done for three types of the background sound speed perturbations: (i) shallow, (ii) deep, and (iii) two-layer models.

We applied the time-distance analysis to both types of simulations and then performed the helioseismic inversion of the obtained travel times to reconstruct the sound-speed perturbation profiles. The time-distance analysis of simulation results for deep and shallow models show good agreement with the theoretical time shifts calculated from the acoustic ray theory for both cases with and without phase speed filtering. The travel time variations in the perturbed regions are negative everywhere and do not exhibit sign inversion at short distances as was reported in \citet{bir09} for the acoustic holography measurements.

The helioseismic inversion of the travel times for the sound-speed show good agreement with the background model profiles for all three cases and for both types of artificial data, obtained with the Cartesian and spherical codes. The inversion procedure produces smoothed results as one can expect, resulting in overestimation of the sound speed perturbation in the shallow case near relatively sharp bottom boundary of the background perturbation. For the two-layer model the smoothing nature of the inversion procedure results in the less steep gradient of the sound speed perturbation near the depth where the perturbation changes sign. These systematic effects should be taken into account in similar analysis of the solar oscillation data.

We summarize our conclusions as follows.
\begin{itemize}
\item
The measurements of acoustic travel times from numerical simulations
of solar oscillation data using the time-distance helioseismology
technique clearly demonstrate that the travel time shifts caused
by subsurface sound-speed perturbations can be well inverted by using the acoustic ray approximation.
\item
The test measurements with and without the phase-speed filtering procedure
demonstrate that this procedure used to improve the signal-to-noise ratio
does not introduce significant systematic errors.
\item
Time-distance analysis and inversions of travel times, obtained from simulations performed by different codes show that the results are not simulation dependant. The sound-speed inversion results show good agreement with the background sound-speed profiles in all cases.
\end{itemize}

\section{Acknowledgments}
This work was partially supported by NASA Living With a Star TR\&T program. The numerical simulations were carried out on Columbia supercomputers at NASA Ames Research Center.

\end{document}